\documentclass[reprint,showpacs,preprintnumbers,amsmath,amssymb,aip]{revtex4-1}
\usepackage{graphicx,wasysym}
\usepackage{dcolumn}
\usepackage{bm}
\usepackage{bbm}
\usepackage{soul}
\usepackage{tabulary}
\usepackage{color}

\newcommand{\tobs}{t_N}
\newcommand{\tint}{\Delta t}
\newcommand{\Nw}{N_\mathrm{w}}

\usepackage{mathrsfs}
\usepackage{natbib}
\usepackage{natmove}
\begin{document}

\preprint{1}

\title{Importance sampling large deviations in nonequilibrium steady states: 1}
\author{Ushnish Ray}
\email{uray@caltech.edu}
\affiliation{%
Division of Chemistry and Chemical Engineering, California Institute of Technology, Pasadena, CA 91125
}
\author{Garnet Kin-Lic Chan}
\email{garnetc@caltech.edu}
\affiliation{%
Division of Chemistry and Chemical Engineering, California Institute of Technology, Pasadena, CA 91125
}

\author{David T. Limmer}

\email{dlimmer@berkeley.edu}
\affiliation{%
Department of Chemistry, University of California, Berkeley, CA 94609
}%
\affiliation{%
Kavli Energy NanoScience Institute, Berkeley, CA 94609
}%
\affiliation{%
Materials Science Division, Lawrence Berkeley National Laboratory, Berkeley, CA 94609
}%

\date{\today}
\begin{abstract}
Large deviation functions contain information on the stability and response of systems driven into nonequilibrium steady states, and in such a way are similar to free energies for systems at equilibrium. As with equilibrium free energies, evaluating large deviation functions numerically for all but the simplest systems is difficult, because by construction they depend on exponentially rare events. In this first paper of a series, we evaluate different trajectory-based sampling methods capable of computing large deviation functions of time integrated observables within nonequilibrium steady states. We illustrate some convergence criteria and best practices using a number of different models, including a biased Brownian walker, a driven lattice gas, and a model of self-assembly. We show how two popular methods for sampling trajectory ensembles, transition path sampling and diffusion Monte Carlo, suffer from exponentially diverging correlations in trajectory space as a function of the bias parameter when estimating large deviation functions.
Improving the efficiencies of these algorithms requires introducing guiding functions for the trajectories.
\end{abstract}


\maketitle

\section{Introduction}

Free energy calculations play a central role in molecular simulations of chemical, biological and material systems. As the governing function for stability, free energies are a central quantity of statistical physics. With the exception of special cases, the evaluation of free energies requires numerical computation. The development and ease of implementation of methods such as umbrella sampling, metadynamics, and Wang-Landau sampling\cite{Torrie1977,Laio2002,Wang2001} has made the computation of equilibrium free energies textbook material\cite{chipot2007free}. With these and related techniques, free energy calculations of realistic models of matter are routinely used to compute phase diagrams, to infer thermodynamic response, and to estimate rates of rare events\cite{allen1989computer, Frenkel2001,landau2014guide}. However, such calculations are traditionally limited to instances where systems are at thermodynamic equilibrium, as most algorithms are typically formulated with an implicit reliance on an underlying Boltzmann distribution of configurations. Large deviation theory provides a theoretical framework for extending free energy calculations to systems away from equilibrium, with large deviation functions acting as the analog of free energies for such systems\cite{Touchette2009}. Though not identical to a thermodynamic potential, large deviation function nevertheless encode stability and often response of nonequilibiurm system. While methods exist to evaluate large deviation functions numerically\cite{merolle2005space,valeriani2007computing,Giardin2006,tchernookov2010list,nemoto2014computation}, few studies on their performance\cite{nemoto2017finite,hidalgo2016finite} or extensions have been reported\cite{Takahiro2016}. Here we evaluate the convergence of two methods, transition path sampling\cite{bolhuis2002transition} and diffusion Monte Carlo\cite{giardina2011simulating}. We find that each samples the same distribution of trajectories and can in principle converge large deviation functions. However, both methods suffer from sampling deficiencies that make applications to high dimensional systems difficult. By illustrating these properties, we provide a basis for what is necessary to systematically extend them for use in molecular models of nonequilibrium steady states. 

Large deviation theory underlies much of the recent progress in nonequilibrium statistical mechanics. Within its context, strikingly simple and general properties for systems driven away from thermal equilibrium have been derived. These include relationships between far-from-equilibrium averages and equilibrium free energies, like the Jarzynski equality and Crooks fluctuation theorem, and symmetries within nonequilibrium ensembles, like the Lebowitz-Spohn and Gallovoti-Cohen symmetries\cite{Jarzynski1997, Crooks1999, Lebowitz1999, GallavottiCohen1995}.  Large deviation functions are cumulant generating functions for the fluctuations of quantities in arbitrary physical systems. Many physical systems of fundamental and applied interest operate away from thermal equilibrium, either because boundary conditions generate a flux of material or energy, or because the individual constituents take in energy to move or exert force. Therefore computing large deviation functions as a means of characterizing the likelihood of fluctuations about a nonequilibrium state has the potential to broaden our understanding of physical processes ranging from crystal growth and molecular self-assembly, to 
 nanoscale fluid flows and instabilities, to the organization of living systems and other active matter \cite{Auer2000, Auer2002, Ren2005, Berhanu2007, Bouchet2009, Easterling2000,weber2013emergence}. 

Large deviation functions are averages taken over a trajectory ensemble. As such, to compute large deviation functions for high dimensional systems, methods for sampling trajectory ensembles must be constructed and employed. Beginning with transition path sampling\cite{Chandler1998}, and followed by its generalizations, such as transition interface sampling\cite{van2005elaborating}, the finite temperature string method\cite{weinan2005transition} such tools have been developed. As will be elaborated on below, such methods sample trajectory space by a sequential random update, resulting in a Monte Carlo walk that uniformly samples trajectories. Similarly, methods that advance an ensemble of trajectories in parallel such as forward flux sampling\cite{allen2009forward}, and the cloning algorithm\cite{giardina2011simulating}, have been established. These methods also sample a trajectory ensemble, though they do so by directly, representing the ensemble by a collection of trajectories that are simultaneously updated in time. Such methods also sample a trajectory ensemble uniformly, and as with sequential update Monte Carlo methods, admit the addition of importance sampling to compute large deviation functions. 

Overwhelmingly these methods have been applied to sample rare trajectories evolved with detailed balance dynamics\cite{bolhuis2015practical}. Notably, such studies have found much success in the study of the glass transition\cite{hedges2009dynamic,speck2012constrained,limmer2014theory} and protein folding\cite{weber2013emergence,mey2014rare}. There have been some efforts to extend trajectory based importance sampling to driven systems, but these have been largely confined to the calculation of simpler observables, rather than large deviation functions\cite{crooks2001efficient,gingrich2016near,allen2005sampling}.  The few applications of path sampling methods to compute 
 large deviations have been limited either to processes occurring with moderately high probability within the nonequilibrium steady state, or to low dimensional model systems\cite{hurtado2010large,speck2011space,hurtado2014thermodynamics}. This is because the statistical efficiency of the standard algorithms becomes exponentially small with the exponentially rare fluctuations needed to be sampled. This has made the computational effort to converge large deviation functions intractable for many realistic systems. As will be illustrated below, the origin of the low sampling efficiency is the small overlap between the distribution of trajectories generated with a model's dynamics, and the distribution whose dynamical fluctuations most contribute to the large deviation function at a given bias. This difficulty manifests itself differently in various methods, with either exponentially small acceptance probability or exponentially large variance in the weights. The generality points to a need to develop robust importance sampling schemes on top of these trajectory based Monte Carlo procedures.  

In the following, we outline the basic theory of large deviations of physical processes, and how they are estimated stochastically with path sampling. This serves to clarify how disparate methods like transition path sampling and diffusion Monte Carlo sample the same trajectory spaces in essentially the same way. We then explore the convergence of standard implementations of different methods by studying systems of increasing complexity: a biased Brownian particle on a periodic potential, a 1d driven lattice gas with volume exclusion, and a 2d model of self-assembly. We use these models to illuminate convergence criteria and the origin of statistical inefficiencies. Throughout we focus on the commonalities between two canonical Monte Carlo methods and aim to draw general conclusions about their efficiencies and how they can be extended to complex systems. These lessons serve to motivate the formulation of a hierarchy of methods to mitigate the low sampling efficiencies by introducing an auxiliary dynamics. The formalism follows the technique of using a trial wavefunction in diffusion Monte Carlo and is discussed in Part II of this work\cite{ray2017importance2}. 
  
\section{Theory and Monte Carlo algorithms}

For compactness and ease of notation, we focus on large deviation functions for processes described by time-independent, Markovian stochastic dynamics. While extensions of large deviation theory to deterministic, non-Markovian dynamics, and time-dependent processes exist, these generalizations will not be considered here. Similarly, we will only consider ensembles of trajectories with a fixed observation time, though generalizations to fixed number of events and fluctuating observation times are straightforward\cite{budini2014fluctuating}. Much of the theory of large deviations summarized here is expanded upon in Ref.~\onlinecite{Touchette2009}. 

In general, Markovian processes obey a continuity equation of the form, 
\begin{equation}\label{Eq:Master}
\partial_t \rho(\mathcal{C}_t,t)= \mathcal{W} \rho(\mathcal{C}_t,t)
\end{equation}
where $\rho(\mathcal{C}_t,t)$ is the probability of observing a configuration of the system, $\mathcal{C}$, at a time $t$ and $\mathcal{W}$ is the linear operator in that configurational space that propagates trajectories. If $\mathcal{C}$ spans a discrete space, then $\mathcal{W}$ is a transition rate matrix and Eq. \ref{Eq:Master} a master equation. If $\mathcal{C}$  spans a continuous space, then $\mathcal{W}$ is a Liouvillian and Eq. \ref{Eq:Master} a generalized Fokker-Planck equation. In either case, provided $\mathcal{W}$ is irreducible, Eq.~\ref{Eq:Master} generates a unique steady state in the long time limit, which in general produces non-vanishing currents and whose configurations do not follow Boltzmann statistics or some generalization.  Only in instances where $\mathcal{W}$ obeys detailed balance is the steady state an equilibrium distribution in which all currents vanish~\cite{van1992}. 

For general nonequilibrium steady states, there exist two generic classes of time extensive observables: \emph{configuration type} observables that are sums of functions of individual configurations,
\begin{equation}
\hat{\mathcal{O}}_a = \sum_{t=1}^{t_N}  \, o_a(\mathcal{C}_t) \, ,
\end{equation}
where $o_a$ is an arbitrary function of a single configuration and \emph{current type} observables that are sums of functions of changes in configurations,
\begin{equation}
\hat{\mathcal{O}}_b = \sum_{t=1}^{t_N} \, o_b(\mathcal{C}_{t-1}\rightarrow \mathcal{C}_{t}) \,,
\end{equation}
where $o_b$ is an arbitrary function of configurations at adjacent times. Within a nonequilibrium steady state, either observable type has an associated probability distribution that can be constructed as a marginal distribution over an ensemble of trajectories. For example, for an observable that is a combination of configurational and current types, $\mathcal{O} = \mathcal{O}_a +\mathcal{O}_b$, 
\begin{equation}
\begin{split}
&p[\mathcal{O}]  = \left < \delta \left [ \mathcal{O} - \sum_{t=0}^{t_N}  \, o_a(\mathcal{C}_t)+ \sum_{t=1}^{t_N} o_b(\mathcal{C}_{t-1}\rightarrow \mathcal{C}_{t}) \right ] \right > \\
&= \sum_{\mathscr{C}(t_N)} P[\mathscr{C}(t_N)] \delta \left [ \mathcal{O} - \sum_{t=0}^{t_N}  \,o_a(\mathcal{C}_t)+ \sum_{t=1}^{t_N} o_b(\mathcal{C}_{t-1}\rightarrow \mathcal{C}_{t}) \right ] \, ,
\end{split}
\label{eq:maineq4}
\end{equation}
where $\langle \dots \rangle$ denotes path ensemble average within the steady state distribution generated by the systems dynamics and $\delta(x)$ is Dirac's delta function. A member of the trajectory ensemble is denoted $\mathscr{C}(t_N)=\{\mathcal{C}_0,\mathcal{C}_1,\dots,\mathcal{C}_{t_N}\}$, which is a vector of all of the configurations that a system has visited over an observation time, $\tobs$. Generically the probability for observing a given trajectory, $\mathscr{C}(t_N)$, is given by 
\begin{equation}\label{Eq:PofC}
P[\mathscr{C}(\tobs)] = \rho(\mathcal{C}_0)\prod_{t=1}^{\tobs} u(\mathcal{C}_{t-1}\rightarrow \mathcal{C}_{t})\, ,
\end{equation}
where $\rho(\mathcal{C}_0)$ represents the distribution of initial conditions and the suppression of an explicit time argument denotes a stationary distribution, and $u(\dots)$ are the transition probabilities. For a given nonequilibrium steady state, $\rho(\mathcal{C}_0)$ is the stationary distribution and the transition probabilities are encoded in the propagator, $\mathcal{W}$.

Observables that are correlated over a finite amount of time admit an asymptotic, time intensive form,  for their distribution function, 
\begin{equation}
\label{Eq:LDRF}
t_N^{-1} \ln p[\mathcal{O}]  \sim -  \phi(\mathcal{O}/t_N) \, ,
\end{equation}
where $\phi(\mathcal{O}/t_N)$ is the rate function or generalized entropy, and $\sim$ denotes the long time limit that is an equality up to sub-extensive terms $\propto 1/t_N$. In this limit, a generating function can be computed by taking the Laplace transform of Eq.~\ref{eq:maineq4}.
\begin{equation}
 \label{Eq:LDF}
\langle  e^{-\lambda \mathcal{O}}\rangle =  \sum_{\mathscr{C}(t_N)} P[\mathscr{C}(t_N)] e^{-\lambda \mathcal{O}} \sim e^{\psi(\lambda) t_N} \, ,
\end{equation}
where $\psi(\lambda)$ is known as the large deviation function or generalized free energy, and $\lambda$ is a counting field conjugate to the observable. Like an equilibrium free energy, the large deviation function is a cumulant generating function and by taking derivatives with respect to $\lambda$ 
\begin{equation}\label{Eq:Mean}
\partial^n \psi(\lambda)/\partial \lambda^n \big |_{\lambda=0} = (-1)^n \frac{1}{\tobs} \langle (\delta \mathcal{O})^n \rangle \, ,
\end{equation}
the $n$th scaled cumulant of $\mathcal{O}$ is generated. Unlike in equilibrium, in general $\lambda$ is not related to a physical field, like temperature.  When the asymptotic limits exists and the rate function is a smooth convex function, then the rate function and large deviation function are related through a Legendre transform,
\begin{equation}\label{Eq:LDF}
 \psi(\lambda) = \inf_\mathcal{O} [\phi(\mathcal{O}/t_N)+\lambda  \mathcal{O}/t_N]
\end{equation}
where $\inf$ is the largest lower bound taken over all $\mathcal{O}$. This structure is reminiscent of thermodynamics, where Legendre transforms change ensembles between conjugate variables. Like in equilibrium thermodynamics, when the large deviation function is non-analytic, or nonconvex, this transform loses information about the fluctuations of the system, as it will return the convex hull of $\phi(\mathcal{O}/t_N)$. Away from these points, when the rate function is convex, the long time limit ensures an equivalence of ensembles\cite{chetrite2015nonequilibrium} between those conditioned on a value of $\mathcal{O}$ and those with fixed field $\lambda$ resulting an average value of $\mathcal{O}$ as in Eq.~\ref{Eq:Mean}. This is analogous to equilibrium in the thermodynamic limit away from points of phase transitions\cite{chandler1987introduction}. 

As an exponential average, estimating large deviation functions through straightforward simulations is difficult for large $\lambda$. Stochastic algorithms have been designed to solve this problem by means of Monte Carlo sampling. In equilibrium free energy calculations, difficulties associated with exponentially rare fluctuations are overcome by introducing a sampling bias in the form of additional terms in the Hamiltonian. Provided the additional terms, the system can be simulated or sampled and the bias can be corrected for easily, because the distribution function is the known Boltzmann form. This basic procedure is the basis of numerous methods such as umbrella sampling, metadynamics and Wang-Landau sampling~\cite{Torrie1977,Laio2002,Wang2001}. 

However, equilibrium importance sampling methods do not translate straightforwardly to nonequilbrium steady states, as the distribution of configurations is not known generically. Added biases in the form of terms in the Hamiltonian, or propagator, are not easily reweighted. However, provided a means of sampling trajectories uniformly from their steady state distribution, then added biases in the evolution of the system can be constructed to push the system  into rare regions and their effects exactly corrected in order to return unbiased estimates of those rare fluctuations.   To importance sample large deviation functions in nonequilibrium steady states, the trajectory ensemble is biased, or tilted, to the form,
\begin{equation}\label{Eq:Pbeta}
P_\lambda[\mathscr{C}(t_N)] = P[\mathscr{C}(t_N)]e^{-\lambda \mathcal{O}[\mathscr{C}(t_N)]-\psi(\lambda) t_N} \, ,
\end{equation}
where, the large deviation function $\psi(\lambda)$ is the normalization constant computable as in Eq.~\ref{Eq:LDF}. The new distribution of trajectories, $P_\lambda[\mathscr{C}(t_N)]$, represent those that contribute most to the large deviation function at a given value of $\lambda$. Ensemble averages for arbitrary observables, $\mathcal{O}'$, within the unbiased distribution and the tilted one, denoted $\langle \dots \rangle_\lambda$, are related by
\begin{equation}
\label{Eq:BiasedEn}
\langle  \mathcal{O}'[\mathscr{C}(\tobs)] \rangle_\lambda = \frac{ \langle \mathcal{O}'[\mathscr{C}(t_N)]  e^{-\lambda \mathcal{O}[\mathscr{C}(t_N)]} \rangle}{\langle e^{-\lambda \mathcal{O}[\mathscr{C}(t_N)]} \rangle} \, ,
\end{equation}
where the denominator is $\exp[\psi(\lambda) t_N]$. In any stochastic sampling method, the ensemble average is estimated by a finite sum of trajectories. Differences between methods dictate how the ensemble of trajectories is constructed, and their efficiencies are determined by how well correlations between trajectories can be engineered and how easily biases or other constraints can be added and removed. In this manner, the importance sampling is not that different from equilibrium free energy calculations, with the exception that the simplifying forms to correct for the added bias provided by the Boltzmann distribution are replaced by implicit forms that must be computed along each trajectory. Below we discuss two Monte Carlo methods for sampling trajectories from nonequilibrium steady states with the correct weights, to which additional importance sampling can be added in order to efficiently evaluate large deviation and rate functions. 

\subsection{Transition path sampling}
The distribution defined by Eq. ~\ref{Eq:PofC} can be sampled using a Monte Carlo algorithm, analogous to path integral quantum Monte Carlo. Chandler and coworkers first identified this analogy and developed transition path sampling (TPS) to importance sample dynamical events occurring in generic classical systems\cite{Chandler1998}. Since its inception, TPS has become an important tool for investigating rare events, typically of the form of barrier crossings in systems evolving around an equilibrium\cite{bolhuis2002transition}. In recent years, it has been used to compute large deviation functions for systems evolving with detailed balance preserving dynamics. Most notably this has been done for model glass-formers\cite{hedges2009dynamic,speck2012first,limmer2014theory} and glassy polymers \cite{weber2013emergence,mey2014rare}. Only a very few studies have employed it in the computation of large deviation functions for driven systems, or nonequilibrium steady states\cite{gingrich2016near,speck2011space}. Nonequilibrium systems constrain the types of Monte Carlo moves that can be easily accommodated, due to incomplete knowledge of the steady state distribution. Below we review the basics of TPS and how it can be used to compute large deviation functions.

TPS generates an unbiased ensemble of trajectories by a sequential random update to an existing trajectory, using a variety of different update moves, and accepting or rejecting each move according to a Metropolis rule. Usually, the update rules are collective, and result in the simultaneous change of many configurations along the trajectory. To derive the Metropolis acceptance rules, the targeted stationary distribution must be known. In order to sample exponentially rare fluctuations of an observable, $\mathcal{O}$, and obtain a reasonable estimate of the large deviation function, trajectories are weighted in proportion to the path ensemble defined in Eq.~\ref{Eq:Pbeta}. Given this distribution, the acceptance criterion for updating the MC walk between two trajectories $\mathscr{C}_\mathrm{o}(\tobs)$ and $\mathscr{C}_\mathrm{n}(\tobs)$ can be determined by the steady state criterion,
\begin{equation}
\begin{split}
P_\lambda[\mathscr{C}_\mathrm{o}] \Gamma[\mathscr{C}_\mathrm{o}\rightarrow \mathscr{C}_\mathrm{n}] \mathrm{Ac}[ \mathscr{C}_\mathrm{o}\rightarrow \mathscr{C}_\mathrm{n}] = 
\\
P_\lambda[\mathscr{C}_\mathrm{n}] \Gamma[\mathscr{C}_\mathrm{n}\rightarrow \mathscr{C}_\mathrm{o}] \mathrm{Ac}[\mathscr{C}_\mathrm{n}\rightarrow \mathscr{C}_\mathrm{o}]
\end{split}
\end{equation}
where $\mathrm{Ac}[\dots]$ is the acceptance probability and $\Gamma[\dots]$ is the generation probability and we have suppressed the arguments of the trajectories, assuming that they are all $\tobs$ in length. Rearranging, we can put the acceptance probability in a Metropolis form: 
\begin{equation}
\label{Eq:TPSaccep}
\mathrm{Ac}[ \mathscr{C}_\mathrm{o}\rightarrow \mathscr{C}_\mathrm{n}] = \mathrm{Min}\left ( 1,e^{- \lambda (\mathcal{O}[
\mathscr{C}_\mathrm{n}] -\mathcal{O}[
\mathscr{C}_\mathrm{o}] ) -\Omega} \right )
\end{equation}
where the first term proportional to $\lambda$ is the bias away from typical events and
\begin{equation}\label{Eq:Balance}
\Omega = -\ln \frac{P[\mathscr{C}_\mathrm{n}] \Gamma[\mathscr{C}_\mathrm{n}\rightarrow \mathscr{C}_\mathrm{o}]}{P[\mathscr{C}_\mathrm{o}] \Gamma[\mathscr{C}_\mathrm{o}\rightarrow \mathscr{C}_\mathrm{n}]} \, 
\end{equation}
depends on how new trajectories are generated. If they are generated using physical dynamics, independent of $\lambda$, then $\Omega$ is a measure of the entropy production to transform between the two trajectories\cite{gingrich2015preserving}. The specific form of the entropy production depends on the details of the Monte Carlo move.

\begin{figure}[t]
\begin{center}
\includegraphics[width=8.5cm]{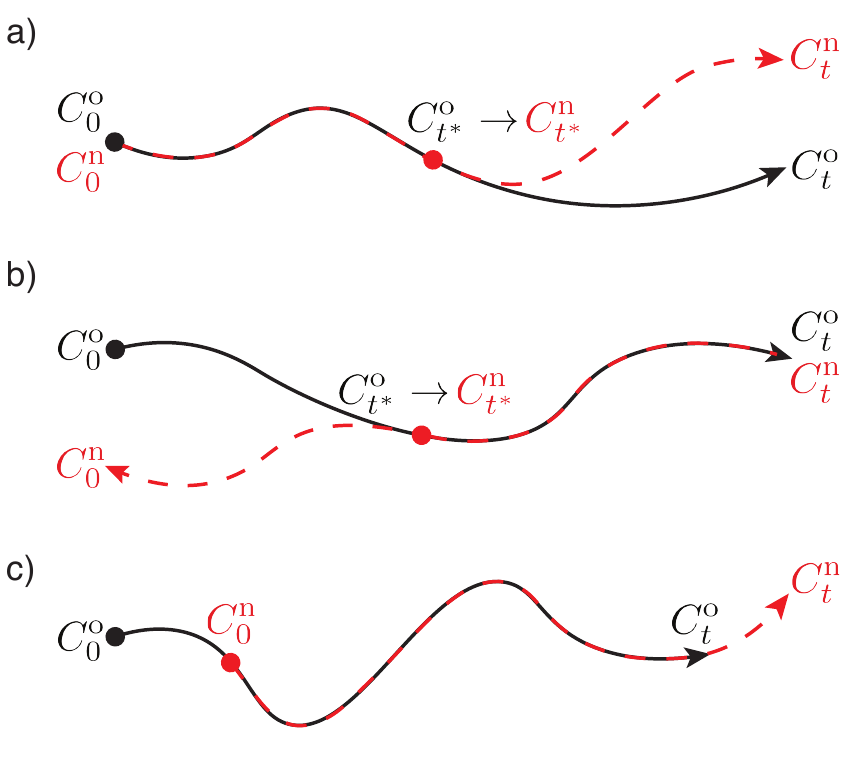}
\caption{Illustration of trajectory update moves used in TPS: a) forward shooting, b) backward shooting, and c) forward shifting.}
\label{Fi:1}
\end{center} 
\end{figure}

For a so-called forward shooting move, a point is picked at random along the trajectory,  $\mathcal{C}_{t^*}$, and the system is reintegrated. This is illustrated  in Fig. \ref{Fi:1}a. The acceptance can be computed by expanding Eq. ~\ref{Eq:Balance},
\begin{equation}
\begin{split}
&e^{-\Omega} = \\
&\frac{\prod_{t=t^*+1}^{\tobs} u(\mathcal{C}^\mathrm{n}_{t-1}\rightarrow \mathcal{C}^\mathrm{n}_{t})
\left (g[\mathcal{C}^\mathrm{n}_{t^*}\rightarrow \mathcal{C}^\mathrm{o}_{t^*}]
\prod_{t=t^*+1}^{\tobs} u[\mathcal{C}^\mathrm{o}_{t-1}\rightarrow \mathcal{C}^\mathrm{o}_{t}]\right )}
 {\prod_{t=t^*+1}^{\tobs} u[\mathcal{C}^\mathrm{o}_{t-1}\rightarrow \mathcal{C}^\mathrm{o}_{t}]
\left (g[\mathcal{C}^\mathrm{o}_{t^*}\rightarrow \mathcal{C}^\mathrm{n}_{t^*}]
\prod_{t=t^*+1}^{\tobs} u[\mathcal{C}^\mathrm{n}_{t-1}\rightarrow \mathcal{C}^\mathrm{n}_{t}]\right )} \\
& \quad\quad =\frac{
g[\mathcal{C}^\mathrm{n}_{t^*}\rightarrow \mathcal{C}^\mathrm{o}_{t^*}]
 }
 {g[\mathcal{C}^\mathrm{o}_{t^*}\rightarrow \mathcal{C}^\mathrm{n}_{t^*}]} 
 = 1
 \end{split}
\end{equation}
where transition probabilities up to $t^*$ trivially cancel in the numerator and denominator because those pieces of the trajectory are the same. The last equality follows from assuming that the shooting point is changed symmetrically with probability $g[\mathcal{C}^\mathrm{n}_{t^*}\rightarrow \mathcal{C}^\mathrm{o}_{t^*}]$, independent of the configuration, for example by drawing new random noise for the stochastic integration. With such a procedure, the move is accepted in proportion to the ratio of $\exp(-\lambda \mathcal{O})$ between the new and old trajectory.

In order to have an ergodic Markov Chain, the entire trajectory from the the initial condition must be sampled.  Unlike in equilibrium, for dynamics that do not obey detailed balance this is problematic, as in general the acceptance probability for backwards moves cannot be manipulated to yield unity even for $\lambda=0$.  For a backwards shooting move, illustrated in Fig.~\ref{Fi:1}b, the acceptance criterion is given by
\begin{equation}
\begin{split}
&e^{-\Omega} = \\
&\frac{
\left (\rho(\mathcal{C}^\mathrm{n}_0) \prod_{t=1}^{t^*} u[\mathcal{C}^\mathrm{n}_{t-1}\rightarrow \mathcal{C}^\mathrm{n}_{t}]\right )
\left (\prod_{t=t^*+1}^{\tobs} \bar{u}[\tilde{\mathcal{C}}^\mathrm{o}_{t-1}\rightarrow \tilde{\mathcal{C}}^\mathrm{o}_{t}]\right )
 }
 {
\left (\rho(\mathcal{C}^\mathrm{o}_0) \prod_{t=1}^{t^*} u[\mathcal{C}^\mathrm{o}_{t-1}\rightarrow \mathcal{C}^\mathrm{o}_{t+1}]\right )
\left (\prod_{t=t^*+1}^{\tobs} \bar{u}[\tilde{\mathcal{C}}^\mathrm{n}_{t-1}\rightarrow \tilde{\mathcal{C}}^\mathrm{n}_{t}]\right )
  }
   \end{split}
\end{equation}
where $\bar{u}[\dots]$ is the time-reverse transition rate, and $\tilde{\mathcal{C}}_i$ the corresponding time reversed state of the system. From stochastic thermodynamics,\cite{Crooks1999,gingrich2015preserving} the time reverse rates are related to the forward rates by the entropy produced, $\Xi$, 
\begin{equation}
\frac{u[\mathcal{C}^\mathrm{n}_{t-1}\rightarrow \mathcal{C}^\mathrm{n}_{t}]}
{\bar{u}[\tilde{\mathcal{C}}^\mathrm{n}_{t-1}\rightarrow \tilde{\mathcal{C}}^\mathrm{n}_{t}]}
=e^{\Xi^{n}(t-1,t)} \, .
\end{equation}
Substituting this relation into the acceptance probability, it is clear that the acceptance rate is exponentially small and contains two terms: a time extensive entropy production and a boundary term with information on the initial conditions,
\begin{equation}
e^{-\Omega} = e^{-[W^\mathrm{n}(t_0,t^*)-W^\mathrm{o}(t_0,t^*)]} \, ,
\end{equation}
where,
\begin{equation}
 W^{\mathrm{o},\mathrm{n}}(t_0,t^*) = \sum_{t=1}^{t^*} \Xi^{\mathrm{o},n}(t-1,t) - \ln \rho(\mathcal{C}^{\mathrm{o},\mathrm{n}}_0) \, . 
\end{equation}
This ratio is problematic because the relative weights for the initial distribution are often not known. Even in cases where the initial distribution is known, or can be neglected as in a long time limit, for a nonequilibrium steady state the time extensive part means that backwards shooting moves are exponentially unlikely to be accepted, with a rate proportional to the length of the move and the number of driven degrees of freedom in the configuration space. 

Alternatively, the initial condition and early time segments of the trajectory can be updated using forward shifting moves, illustrated in Fig.~\ref{Fi:1}c. In such a move, a configuration from the middle of the trajectory is randomly selected,  $\mathcal{C}_{t^*}$, and moved to the beginning of a new proposed trajectory. In order to conserve the total length of the trajectory, an additional segment is generated from the last configuration of the old trajectory. The acceptance criterion for such a move can be written,
\begin{equation}
e^{-\Omega} = 
\frac{
\rho(\mathcal{C}^\mathrm{n}_{t^*})
\prod_{t=1}^{t^*} \bar{u}[\tilde{\mathcal{C}}^\mathrm{n}_{t}\rightarrow \tilde{\mathcal{C}}^\mathrm{n}_{t-1}]
 }
 {
\rho(\mathcal{C}^\mathrm{o}_0) \prod_{t=1}^{t^*} u[\mathcal{C}^\mathrm{o}_{t-1}\rightarrow \mathcal{C}^\mathrm{o}_{t}]
  }
  =1 \, ,
\end{equation}
where the final equality assumes that the nonequilibrium driving is time-reversal symmetric, and that the system is evolving within a stationary steady state. This cancelation occurs because under these assumptions, the proposal and path weights are generated with the same dynamics. This is analogous to previous work whereby new initial conditions are generated by short trajectories, followed by reintegration of the full path.\cite{crooks2001efficient} Because the dynamics we consider are stochastic and irreducible, forward shooting and forward shifting are sufficient to ensure an ergodic Markov chain, as over repeated TPS moves the entire trajectory is replaced. This is  distinct from other forward only propagating methods like forward flux sampling where the initial conditions are not updated, which can result in biased sampling if sufficiently large initial distributions are not used\cite{van2012dynamical,bolhuis2015practical}. However, trajectories decorrelate more slowly without backwards shooting as many shooting and shifting moves are needed to generate completely new trajectories. The acceptance criterion for backwards shifting can be analogously derived, and presents similar problems as backwards shooting.

The moves outlined above, together with the acceptance criterion of Eq.~\ref{Eq:TPSaccep}, allow TPS to be used to sample trajectory ensembles of arbitrary nonequilibrium systems in proportion to their correct statistical weight. Expectation values of the form of Eq.~\ref{Eq:BiasedEn}, can be estimated straightforwardly by
\begin{equation}
\langle  \mathcal{O}' \rangle_\lambda = \frac{1}{N_\mathrm{traj}} \sum_{i=1}^{N_\mathrm{traj}} \mathcal{O}'[\mathscr{C}_i(\tobs) | \lambda] 
\end{equation}
where $N_\mathrm{traj}$ is the number of trajectories harvested by TPS. This is because by construction they sample $P_\lambda[\mathscr{C}(\tobs)]$ without a bias. Such expectation values can be related to the unbiased steady state distribution using Eq.~\ref{Eq:BiasedEn}. In order to estimate the large deviation or rate function, simulations can then be run with multiple values of $\lambda$, and provided the distributions collected at various values of $\lambda$ overlap, $\psi(\lambda)$ can be computed from the relation,
\begin{equation} 
\label{Eq:Wham}
\phi_\lambda(\mathcal{O}) = \phi(\mathcal{O})  + \lambda \mathcal{O}/ \tobs +  \psi(\lambda)
\end{equation}
where $\phi_\lambda(\mathcal{O})$ is the marginal distribution, or unnormalized rate function, generated from Eq.~\ref{Eq:Pbeta}. For multiple distributions, calculations with many values of $\lambda$ can be related with standard histogram reweighting techniques, such as the weighted histogram analysis method  or the multistate bennet acceptance ratio (MBAR), which provide optimal estimates of $\psi(\lambda)$, and unbiased $\phi(\mathcal{O})$ computed far into the tails of its distribution\cite{kumar1992weighted,shirts2008statistically}. Histogram reweighting techniques can only determine $\psi(\lambda)$ up to a global constant, but because the unbiased dynamics are normalized, $\psi(0)=0$, so the full function of $\psi(\lambda)$ can be uniquely determined.  These procedures are a straightforward generalization of tradition umbrella sampling calculations in equilibrium, only now with an ensemble of trajectories.

\subsection{Diffusion Monte Carlo} \label{subsec:DMC}
An alternative to sampling trajectory space with sequential random updates, is to sample trajectories in parallel. This is the strategy proposed in diffusion Monte Carlo (DMC), and has been adopted for the study of nonequilibrium classical stochastic dynamics by algorithms known as forward flux sampling, and the so-called cloning algorithm \cite{Grassberger2002, Takahiro2016,tchernookov2010list}. In both methods, an ensemble of initial conditions is generated using some prior distribution. Each initial condition, referred to here as a walker, is propagated for some short time, creating an ensemble of short trajectories. Depending on the specific algorithm and observable of interest, importance sampling is incorporated by assigning weights to each walker that accumulate over the short trajectories and upon iterations of parallel updates to the full trajectory ensemble. Generically, walkers are weighted in proportion to the biased ensemble defined in Eq. ~\ref{Eq:Pbeta}. The cloning algorithm offsets the exponential weights with a population dynamics, whereby walkers with low weight are deleted and those with high weight are replicated or ``cloned". The cloning algorithm in particular has been successful in computing large deviation functions for model lattice systems\cite{garrahan2009first,bodineau2012finite}, and some small atomistic systems\cite{pitard2011dynamic}. Mostly it has been used to study models of transport in 1 dimension \cite{hurtado2011spontaneous,mitsudo2011numerical,giardina2011simulating}. Below we review the basics of DMC and how it can be used to compute large deviation functions.

\begin{figure}[t]
\includegraphics[width=8.5cm]{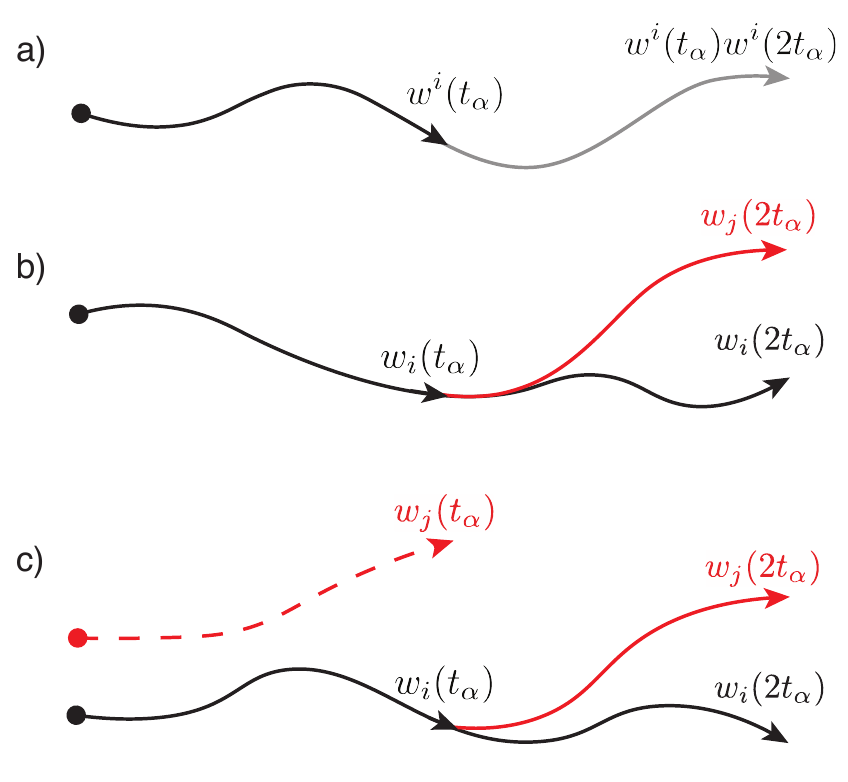}
\caption{Illustration of the diffusion Monte Carlo algorithm. The red and black lines represent different walker trajectories. a) Illustrates the multiplicative weight of trajectories constructed during forward propagation. b) Non-constant walker population method, whereby branching and deletion is done in proportion to $w_i$. c) Constant walker population method where branching and deletion result in continuous (solid lines) and discontinuous (dashed lines) trajectories.}
\label{Fi:cloning}
\end{figure}

The form of $P_\lambda[\mathscr{C}(t)]$ suggests an alternative way in which trajectories may be generated. Expanding $P_\lambda$, we get
\begin{align}
&P_\lambda[\mathscr{C}(\tobs)] = \nonumber\\
&\frac{\rho(\mathcal{C}_0) \prod_{t=1}^{\tobs} u(\mathcal{C}_{t-1}\rightarrow \mathcal{C}_{t}])e^{-\lambda o(\mathcal{C}_{t-1}\rightarrow \mathcal{C}_{t})}} 
{\sum_{\mathscr{C}(\tobs)} \rho(\mathcal{C}_0) \prod_{t=1}^{\tobs} u(\mathcal{C}_{t-1}\rightarrow \mathcal{C}_{t}) e^{-\lambda o(\mathcal{C}_{t-1}\rightarrow \mathcal{C}_{t})}}
\label{dmc:eq2} 
\end{align}
where, we have written $\mathcal{O}$ as a current type observable, $\mathcal{O}[\mathscr{C}(\tobs)]= \sum_{t=1}^{\tobs} o(\mathcal{C}_{t-1}\rightarrow \mathcal{C}_{t})$, otherwise it could be an additive function of single configurations. The argument inside the product is the transition probability times a bias factor. This combination of probabilities does not represent a physical dynamics, as it is unnormalized. However, it can be interpreted as a population dynamics as was done by Giardina, Kurchan, and Peliti and is done in standard DMC, where the nonconservative part proportional to the bias is represented by adding and removing walkers\cite{Giardin2006}. 

In particular, in the cloning algorithm, trajectories are propagated in two steps. First $N_\mathrm{w}$ walkers are propagated according to the normalized dynamics specified by $u[\mathcal{C}_{t-1}\rightarrow \mathcal{C}_{t}]$ for a time, $\tint$. Over this time, the bias is accumulated according to
\begin{equation}
w_i(t,\tint) = \exp \left ( - \lambda O[\mathcal{C}_i(t-\tint),\mathcal{C}_i(t)] \right )
\end{equation}
where $O[\mathcal{C}_{t-\tint},\mathcal{C}_t] = \sum_{t'=t-\tint}^{t} o[\mathcal{C}_{t'}\rightarrow \mathcal{C}_{t'+1}]$ is the collected current, which due to the multiplicative structure of the Markov chain is simply summed in the exponential. After the trajectory integration, $n_i(t)$ identical copies of the $i$th trajectory are generated in proportion to $w_i(t,\tint)$. This is known as a branching step. In standard quantum-DMC, $\Delta t $ is chosen to be the time step  corresponding to trotterization of the Hamiltonian\cite{Reynolds1982}. Additionally, instead of using the bias to branch, it is advantageous to incorporate it into the integration step and use the normalization as the branching weight. However, in many applications of non-equilibrium stationary states calculating the normalization may not be straightforward.

There are different ways of performing the branching step that either hold the total number of walkers constant or not. In a simple formulation with a non-constant walker population, the number of replicated walkers is determined stochastically by
\begin{equation}
\label{Eq:weights1}
n_i(t) = \lfloor w_i(t,\tint) + \xi\rfloor 
\end{equation}
where $\xi$ is a uniform random number between 0 and 1, and $\lfloor \dots \rfloor$ is the floor function. Alternatively, the floor function may be avoided by using a partial weight for the walker, which decreases the statistical noise. The process of integration followed by branching is continued until $t = t_N$. In this case, the large deviation function can be directly evaluated from the walker population as,
\begin{equation}
\psi(\lambda) = \frac{1}{\tobs} \ln \prod_{t=1}^{\tobs/\tint} \frac{\Nw(t)}{\Nw(t-1)}
\end{equation}
which follows from the equivalence between the normalization constant of the biased distribution and the large deviation function. This relation clarifies that the number of walkers needed to resolve $\psi(\lambda)$ is exponentially large in $\lambda$, and the time extensive observable, $\mathcal{O}$. Because of this exponential dependence, this method is not practical and requires that the population is controlled in some way to mitigate walker explosion or total walker deletion. 

One way to do population control is to formulate the DMC algorithm with a constant walker population. In this case, as before each walker is initially integrated over a time $\tint$. At this end of the short trajectory $n_i(t)$ copies of the $i$th walker are generated stochastically by
\begin{equation}
\label{Eq:weights2}
n_i(t) = \left \lfloor N_\mathrm{w} \frac{w_i(t,\tint)}{\sum_{j=1}^{N_\mathrm{w}} w_i(t,\tint)} + \xi \right \rfloor 
\end{equation}
where $\xi$ is a uniform random number between 0 and 1. Generally this process will result in a different population of walkers, and so deficiencies or excesses in the walker population are compensated by cloning or deleting walkers uniformly. In this case, the large deviation function can be evaluated at each time as
\begin{equation}
\psi^t(\lambda) =  \ln \frac{1}{\Nw} \sum_{i=1}^{\Nw} w_i(t)
\end{equation}
which is an exponential average over the bias factors of each walker. 
This local estimate can be improved by averaging over the observation time,
\begin{equation}
\label{Eq:DMCpsi2}
\psi(\lambda) = \frac{1}{\tobs} \sum_{t=1}^{\tobs/\tint} \psi^t(\lambda)
\end{equation}
which upon repeated cycles of integration and population dynamics yields a refined estimate of $\psi(\lambda)$. 
Since the population dynamics represents the path probability, averages can be computed with walker weights equal to
\begin{equation}
\langle  \mathcal{O}' \rangle_\lambda =  \frac{1}{N_\mathrm{w}}\sum_{i=1}^{N_\mathrm{w}} \mathcal{O}'[\mathscr{C}_i(t_N)], 
\end{equation}
where the trajectory $\mathscr{C}(t)$ has been generated according to $P_\lambda[\mathscr{C}(t)]$ in Eq.~\ref{dmc:eq2}. This estimator is only valid if the number of walkers is held constant. We note here that instead of using a constant population of walkers it is possible to use a non-constant walker population with a chemical potential as is done in standard quantum DMC. This can lead to better statistics since in this case the uniform redistribution of excesses or deficit of walkers is avoided\cite{Umrigar1993}. 
 
The cloning algorithm utilizes the multiplicative structure of the Markov dynamics to tune the relative times between population steps, $\tint$, in order to effect adjustments in the size of the biases used to add or remove trajectories. The choice of $\tint$ is critical as it scales the overlap between the exponential bias and the generating dynamics. 
Population sizes can be decreased for smaller $\tint$ since the distribution from the generating dynamics is wider for smaller $\tint$ and the exponential decays more slowly since $\mathcal{O}$ is time-extensive.  However, the wider the distribution, the larger the variance in $\psi^t(\lambda)$, so local estimates of the trajectory distribution and its bias are poorer.
 This tension results in a sensitivity to $\tint$ that requires it to be chosen such that it is large enough to affect good local estimates of the large deviation function, but small enough that the overlap between the bias factor and the bare dynamics is large. 
However, it cannot be too large, since population control is only done after $\tint$, and so the systematic error from finite population will grow with $\tint$ as will the stochastic error from finite walker population. 

The time-series that result from DMC can be considered in two ways: in terms of continuous or discontinuous trajectories as illustrated in Fig.~\ref{Fi:cloning}.  The continuous trajectories can be constructed by tracing the configurations backwards from $\mathscr{C}(\tobs)$ keeping those branches that have been cloned. In this case trajectories sample exactly $P_\lambda[\mathscr{C}(t)]$ and expectation values can be computed just as in TPS. This means that histogram reweighing procedures outlined in the previous section can be used to compute $\phi(\mathcal{O})$, instead of the typical method of computing $\phi(\mathcal{O})$ from an inverse of the Legendre transform. The discontinuous paths correspond to trajectories constructed by tracking the configurations of a walker as they are  propagated forward in time. In this case, each cloning event introduces abrupt change in configuration space since a walker's configuration may be replaced by configurations from other walkers that carry larger weights. The ensemble described by discontinuous trajectories are not equal to the trajectories of the biased ensemble generally. For sufficiently large number of walkers and long enough propagation time the two ensembles are strongly overlapping. 

\subsection{Comparisons between TPS and DMC}

Both TPS and DMC are Monte Carlo procedures to sample trajectory space with the correct path weight.  While their implementations and executions are different, both evaluate expectation values stochastically and do so by updating the system of interest using its unmodified propagator. In both methods, the computational cost scales as the number of trajectories generated times the observation time. Systematic errors arising from finite length trajectories or finite walker populations can be alleviated in both, as will be shown below. The resulting Monte Carlo walk in trajectory space is explored differently in the different algorithms, provided each can relax to a stationary distribution, estimates of comparable statistical accuracy require the same number of uncorrelated trajectories and thus the same computational effort. 

In TPS, the sequential update means that complete trajectories of length $\tobs$ are constructed and so global constraints such as temporal boundary conditions imposed on the trajectory ensembles are naturally accommodated. Indeed, transition path sampling was conceived of as a means of sampling equilibrium trajectory ensembles with hard constraints on the initial and final configurations, as needed in the study of barrier crossing events. However, because large deviation functions result from a long time limit, individual long trajectories must be constructed. Maintaining correlations for such long trajectories with standard shooting and shifting moves is difficult especially when the ensemble sampled by the natural system dynamics does not have significant overlap with the ensemble that contributes to the large deviation function at finite $\lambda$. As will be discussed below, this can result in significant statistical errors. 

In contrast to TPS, DMC constructs paths in parallel using an ensemble of walkers and implicitly benefits from modern computer architectures. However, the propagation in DMC is uni-directional with path construction relying heavily on the distribution of paths in a local interval of time, $\tint$. Accurate sampling of the local distribution therefore becomes exceedingly important and due to the exponential scaling of the bias becomes hard to accomplish with limited walker populations. 
Formally the number of walkers needed to accurately converge the calculation scales exponentially with $\lambda$. This complexity results in individual trajectories carrying exponentially larger weights than others, which means they are likely to be replicated more often. This can produce an ensemble of walkers that becomes highly correlated over long times and finite walker populations. Modulating these walker correlations is the key to controlling statistical errors in DMC. 


\section{Convergence criteria and statistical uncertainty}
In order to understand the utility of these algorithms, and undercover their limitations, we have studied three different models. The first is forced Brownian motion that evolves in continuous space and time, but whose large deviation function can be computed exactly. The second is a driven lattice gas that evolves in continuous time but discrete space,  whose two-body interactions result in correlated many-body dynamics. The third is a minimal model of nonequilibrium self-assembly that evolves in discrete space and time, but exhibits a nontrivial phase diagram including critical fluctuations. These three systems allow us to ground a discussion of the systematic and statistical errors that occur in calculating large deviation functions.\\

\subsection{Long time or large population limit}

Finite time and finite walker population\cite{nemoto2017finite} are the largest source of systematic error in the calculation of large deviation functions. Other sources of error that exist independent of importance sampling, such as those resulting from the discretization of continuous time dynamics are well understood and controllable\cite{Frenkel2001}. The definitions for the limiting forms of the probability distribution or its cumulant generating function, Eq.~\ref{Eq:LDRF} and \ref{Eq:LDF}, are exponential equalities. Finite time effects result from sub-exponential terms whose origins are from finite time correlations or from temporal boundary conditions. In TPS, the long time limit is approached by increasing $\tobs$ directly. Deviations from the asymptotic values of $\phi$ or $\psi$ come directly from the temporal boundary conditions that are free to fluctuate independently, and will  bias initial and final parts of the trajectory towards values typical of the system at $\lambda=0$.  In DMC, provided trajectories have been integrated into the steady state, the trajectory space is rotated into a space of walkers, and the effective long time limit can be approached by increasing walker population at fixed integration time. Because of the assumed ergodicity of the nonequilibrium steady state, averages computed from an ensemble of trajectories are expected to be equivalent to those computed from an infinitely long single trajectory. In DMC, provided the fixed integration time is long compared to the correlation times of the system dynamics, $\phi$ or $\psi$ can be converged solely by increasing the walker population. Alternatively, one can take a hybrid approach, where a finite number of walkers can be compensated by longer propagation times.
 
\begin{figure}[t]
\begin{center}
\includegraphics[width=8.5cm]{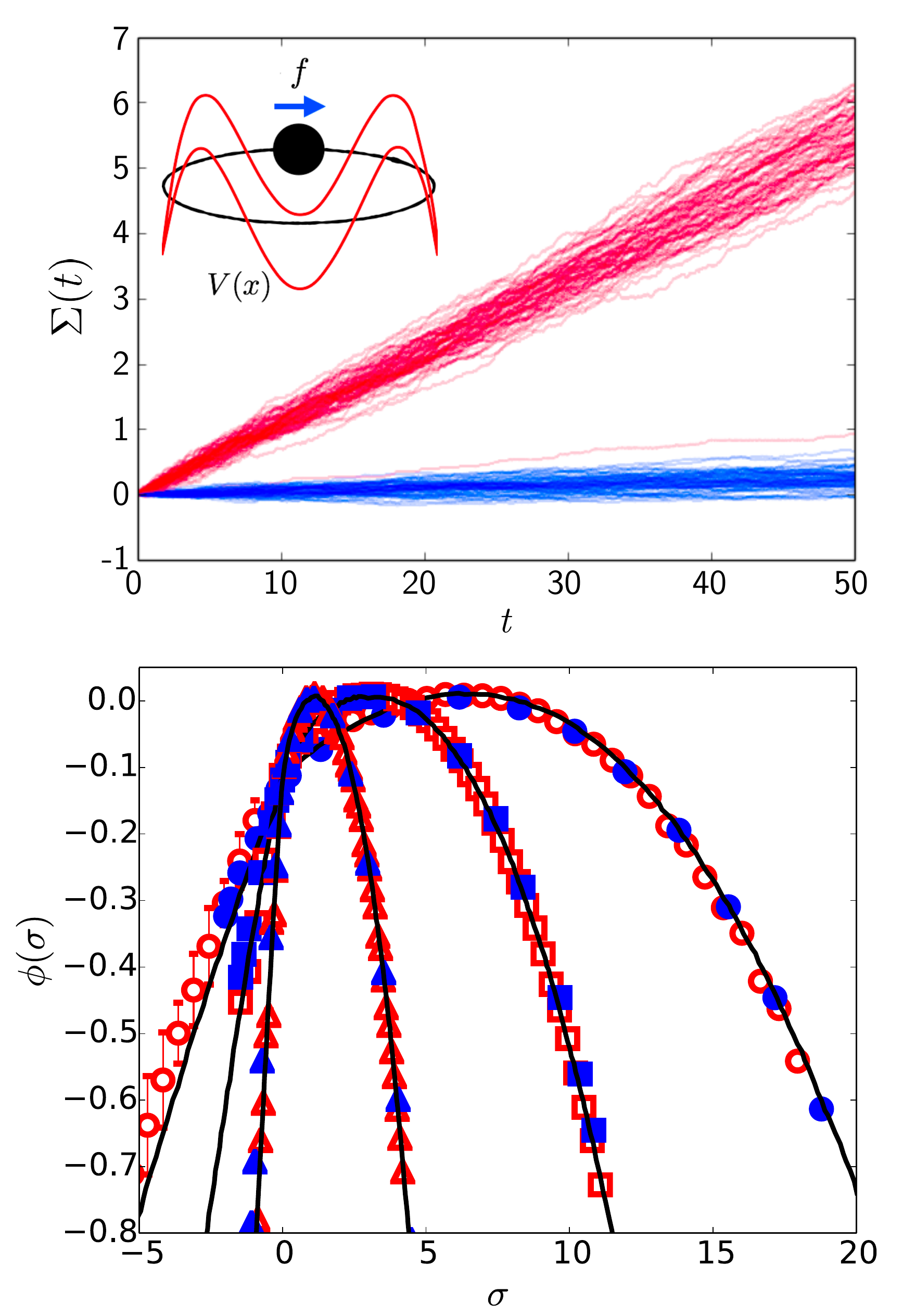}
\caption{Rare entropy production fluctuations of a driven Brownian walker. a) Typical trajectories for instances of large $f^*=1.5$ (red), and small $f^*=0.5$ (blue). Inset is a schematic of the model. b) Rate functions for the entropy production for $f^*=0.8$ (triangles), 1.0 (squares), and 1.5 (circles) computed with TPS (red) and DMC (blue). The black curves are numerically exact results.}
\label{Fi:DBW}
\end{center} 
\end{figure}
To explore the systematic errors arising in the calculations of large deviation functions, we consider a paradigmatic model for driven systems: Brownian motion in a one-dimensional periodic potential with an external force\cite{reimann2001giant,mehl2008large}. The system is sketched in Fig. \ref{Fi:DBW}a. This model has been studied extensively in the context of the fluctuation theorem and is exactly solvable numerically by diagonalization of the Fokker-Planck operator through basis set expansion techniques \cite{Touchette2016}. As such, it provides an ideal reference to understand errors resulting from finite trajectory length or finite walker population.\\ 

The overdamped dynamics of a Brownian particle moving in a closed ring geometry with circumference $L$ and external periodic potential $V(x) = v_o \cos(2\pi x/L)$ obeys a Langevin equation
\begin{equation}
\dot{x} = -\nabla V(x)+ f+\xi \, ,
\end{equation}
where the dot denotes the derivative with respect to $t$. The particle is driven out of equilibrium through the non-gradient force $f$, resulting in a non-vanishing current around the ring. The random force $\xi$, describes the effective interactions between the particle and environment fixed at temperature $T$ with statistics
\begin{equation}
\label{Eq:DBW}
\langle \xi(t) \rangle=0 \quad \langle \xi(t)\xi(t') \rangle = 2\mu T \delta(t-t')
\end{equation}
where the mobility of the particle has been set to unity. In the results we present below, we set $L=T=1$, $v_o=2$ and consider the effects of changing a non-dimensionalized external force, $f^*=f/2\pi v_o\ge 0$.

The dynamics generated by Eq. \ref{Eq:DBW} are well characterized by two limiting behaviors. In the limit that $f^*$ is small, a typical trajectory is localized  near a potential minimum with rare excursions to adjacent minima. Under such conditions, the average particle flux around the ring is small. In the limit that  $f^*$ is large, a typical trajectory exhibits frequent hops between adjacent minima, biased in the direction of the non-gradient force. Under these conditions, the average particle flux around the ring is large. Examples of both types of trajectories are shown in Fig. \ref{Fi:DBW}a. A parameter that characterizes the dynamical behavior of the system is the entropy production
\begin{equation}
\Sigma(t) = \int_0^{t} f \, \mathrm{d}x(t') = f \Delta x(t)
\end{equation}
which is time extensive and in this case simply proportional to the particle flux around the ring.  We denote its time intensive counterpart, $\sigma(\tobs)= \Sigma(\tobs)/\tobs$. In order to evaluate the finite time effects for TPS and DMC, we compute the rate function for entropy production, $\phi(\sigma)$, and its associated large deviation function, $\psi(\lambda)$, for this model. 

To simulate the model of driven Brownian motion, we use a second order stochastic Runga-Kutta integrator with timestep 0.001.\cite{branka1999algorithms} For calculations using TPS we simulated trajectories of length $\tobs=2000$, using forward shooting and shifting moves and correlated noises\cite{gingrich2015preserving}. Trajectories are harvested using the acceptance criterion given in Eq.~\ref{Eq:TPSaccep}, with a bias of $\exp(-\lambda \tobs \sigma(t))$, for $\lambda$ values ranging between -1 and 1 in increments of 0.025. Shooting and shifting points were optimized to result in a 30$\%$ acceptance ratio and around 10$^6$ trajectories were harvested for each value of $\lambda$. Rate functions were computed using MBAR and error bars were estimated using a bootstrapping analysis. For calculations using DMC, we set $\tint=$0.05 and $\tobs=20$, and a constant walker population of $\Nw=$10000. Branching probabilities were computed from Eq.~\ref{Eq:weights2}. Values of $\lambda$ between -1 and 1 were used to compute $\psi(\lambda)$ using Eq.~\ref{Eq:DMCpsi2}, and a numerical Legendre transform  was used to determine $\phi(\sigma)$. 
 
Figure \ref{Fi:DBW}b compares the rate functions for three different cases, characterized by an increasing force, $f^*$. For finite $f^*$, $\phi(\sigma)$ exhibits a minimum at a finite value for the entropy production. With increasing $f^*$, the mean value of $\sigma$ increases, and the curves broaden. In addition, there is an asymmetry which becomes more notable with increasing $f^*$. This latter feature arises from breaking microscopic reversibility and is codified in the fluctuation theorem,
\begin{equation}
\label{Eq:FT}
\phi(\sigma)-\phi(-\sigma) = \sigma \, .
\end{equation}
As derived for diffusion processes by Kurchan\cite{Kurchan1998}, the fluctuation theorem relates the probability of generating entropy to the probability of destroying it. Equation \ref{Eq:FT} implies that as the entropy production gets larger, the likelihood of seeing a fluctuation that reduces entropy becomes exponentially rarer. Figure \ref{Fi:DBW}b compares $\phi(\sigma)$ computed with TPS and DMC to numerically exact results. For all three cases studied, both methods are able to recover the exact result quantitatively. From the fluctuation theorem, Eq. \ref{Eq:FT}, its clear that trajectories that generate negative entropy are rarer, and as a result the error bars for both TPS and DMC are larger for $\sigma<0$. The ability to reproduce the exact result itself is not surprising for this simple model. There are no approximations introduced in the algorithms, so in the limit of well sampled ensembles, the converged expectation values should agree with the exact result within their statistical error bar. It is then only a question of how quickly the  long time or large walker population limit is reached, so that finite time effects are converged to a given accuracy.

\begin{figure}
\begin{center}
\includegraphics[width=8.5cm]{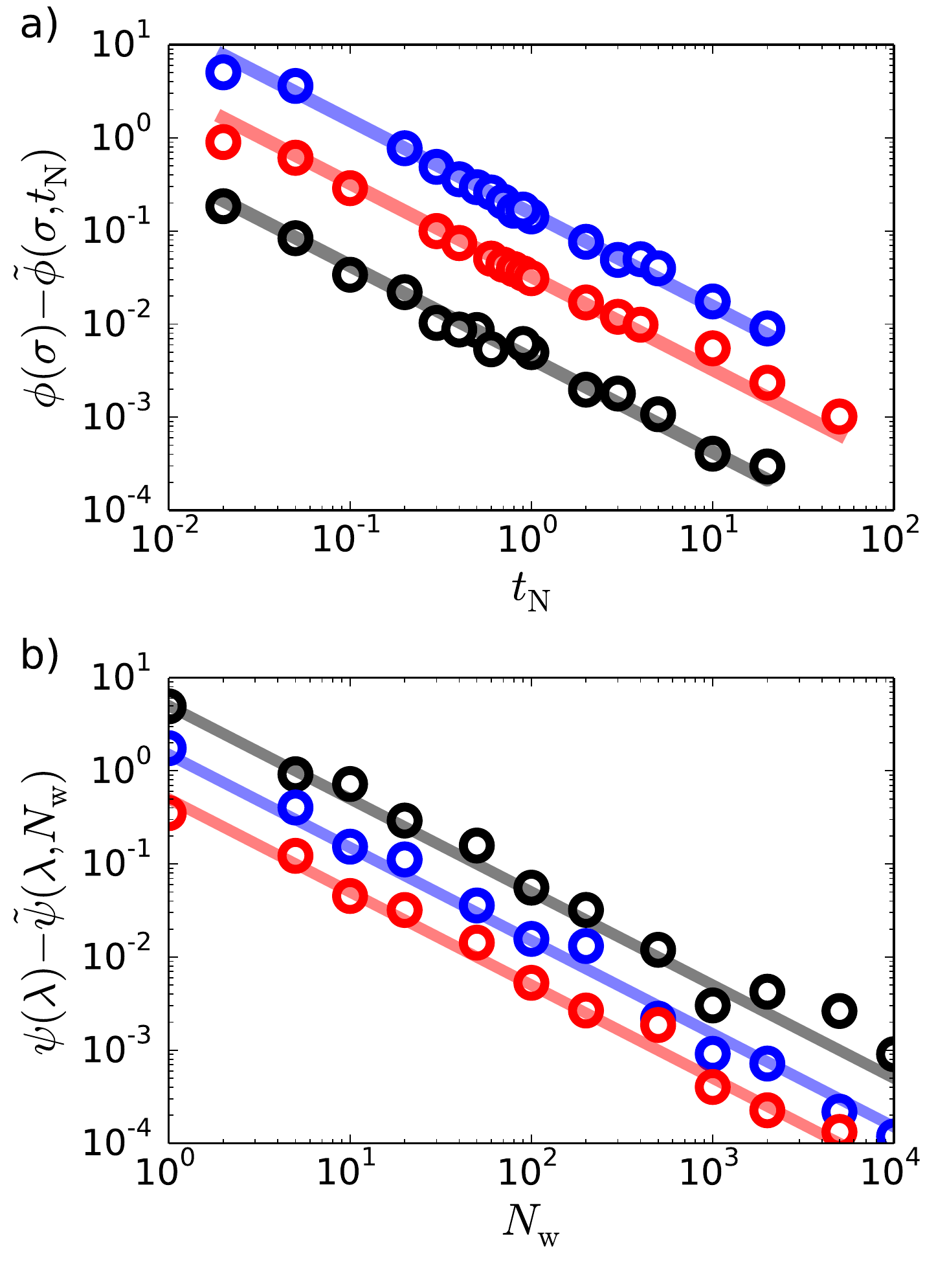}
\caption{Systematic errors in large deviation function of the entropy production for the driven Brownian walker with $f^*=0.8$, $v_o=2$. (a) Finite time error computed with TPS for $\lambda = -0.25$ (blue), 0.0 (red), and 0.25 (black). Solid lines are best fits of the form $A/t_N$. (b) Finite walker number error computed with DMC for $\lambda = -0.5$ (blue), -0.2 (red), and 0.2 (black). Solid lines are best fits of the form $B/\Nw$.}
\label{Fi:4}
\end{center} 
\end{figure}

For the model of driven Brownian motion, we find that the long time limit for the rate function is approached as,
\begin{equation}
\phi(\sigma) - \tilde{\phi}(\sigma,\tobs) \propto 1/\tobs + \mathcal{O}(1/\tobs^2)
\end{equation} 
where $\tilde{\phi}(\sigma,\tobs)$ is the finite time estimate of the asymptotic $\phi(\sigma)$. To leading order, finite time corrections scale as $1/\tobs$ with a prefactor that depends on the correlation time for fluctuations in $\sigma$. Data taken from TPS calculations with varying $\tobs$ are shown in Fig.~\ref{Fi:4}a for a number of different $\lambda$ values. In this model, time correlations arise due to  barrier crossing events that result in large changes in $\sigma$ relative to the typical small amplitude changes from fluctuations of the particle at the bottom of the potential well. For small $f^*$ or $\lambda>0$ these barrier crossing events occur infrequently, and as a consequence the times along a trajectory where they occur are uncorrelated. By contrast, for large $f^*$ or $\lambda<0$, barrier crossings occur often, and accordingly $\sigma$ fluctuations exhibit longer correlation times. Because of this, the prefactor accompanying the $1/\tobs$ correction to the rate function becomes larger for increasing $\lambda$, or increasing $f^*$, as shown in Fig.~\ref{Fi:4}a.

Provided $\tint$ is large enough to produce the steady state, the systematic errors in DMC depend on the number of walkers. Formally, for branching rates obeying Eq.~\ref{Eq:weights2}, the number of walkers required to offset the bias is exponential in $\lambda \sigma \tint$. For a constant walker number algorithm, Nemoto et al. have shown\cite{nemoto2016finite} that the leading order error in the rate function scales as,
\begin{equation}
\psi(\lambda) - \tilde{\psi}(\lambda,\Nw) \propto 1/\Nw + \mathcal{O}(1/\Nw^2)
\end{equation}
where $\psi(\lambda)$ is the Legendre transform of $\phi(\sigma)$ and $\tilde{\psi}(\lambda,\Nw)$ its finite population estimate. This is indeed what we find for the driven Brownian motion model. As shown in Fig.~\ref{Fi:4}b, for various values of $\lambda$ we find systematic errors in the estimate of the large deviation function are inversely proportional to the walker population. 
This is due to the exponential error in representing the steady state distribution made by DMC with a finite number of walkers. As an expectation value of an exponential quantity, $\psi(\lambda)$ consequently depends linearly on this population error. 
The coefficient determining the proportionality between the stochastic error and the number of walkers is related to the number of correlated walkers, which increase with increasing $\lambda$. 

\subsection{Correlations in trajectory space}
In TPS and DMC with branching, expectation values are estimated by summing over trajectories generated with the correct path weights, so that each entry into the sum is weighted equally. Because in practice this sum is finite, the existence of statistical errors determines the accuracy with which expectation values, like large deviation functions, can be computed. For a given number of trajectories, the statistical efficiency is determined by the number of trajectories that are uncorrelated. In TPS, the sequential update results in correlations between trajectories that originate from a common trajectory and share some configurations in common. The relevant measure of these correlations is the time, in number of TPS moves, over which trajectories remain correlated. In DMC, the parallel branching rules replace lower weight walkers with those of higher weight, resulting in a given walker history that can be traced back to some number of common ancestors. Correlations within the trajectory ensemble are measured by the number of walkers that originate from the same source. In both cases, for traditional implementations of these algorithms, we find that correlations rapidly increase for increasingly rare fluctuations. In practice this renders the convergence of large deviation functions untenable for all but the smallest system size, or $\lambda$. 

\begin{figure}[h]
\begin{center}
\includegraphics[width=8.5cm]{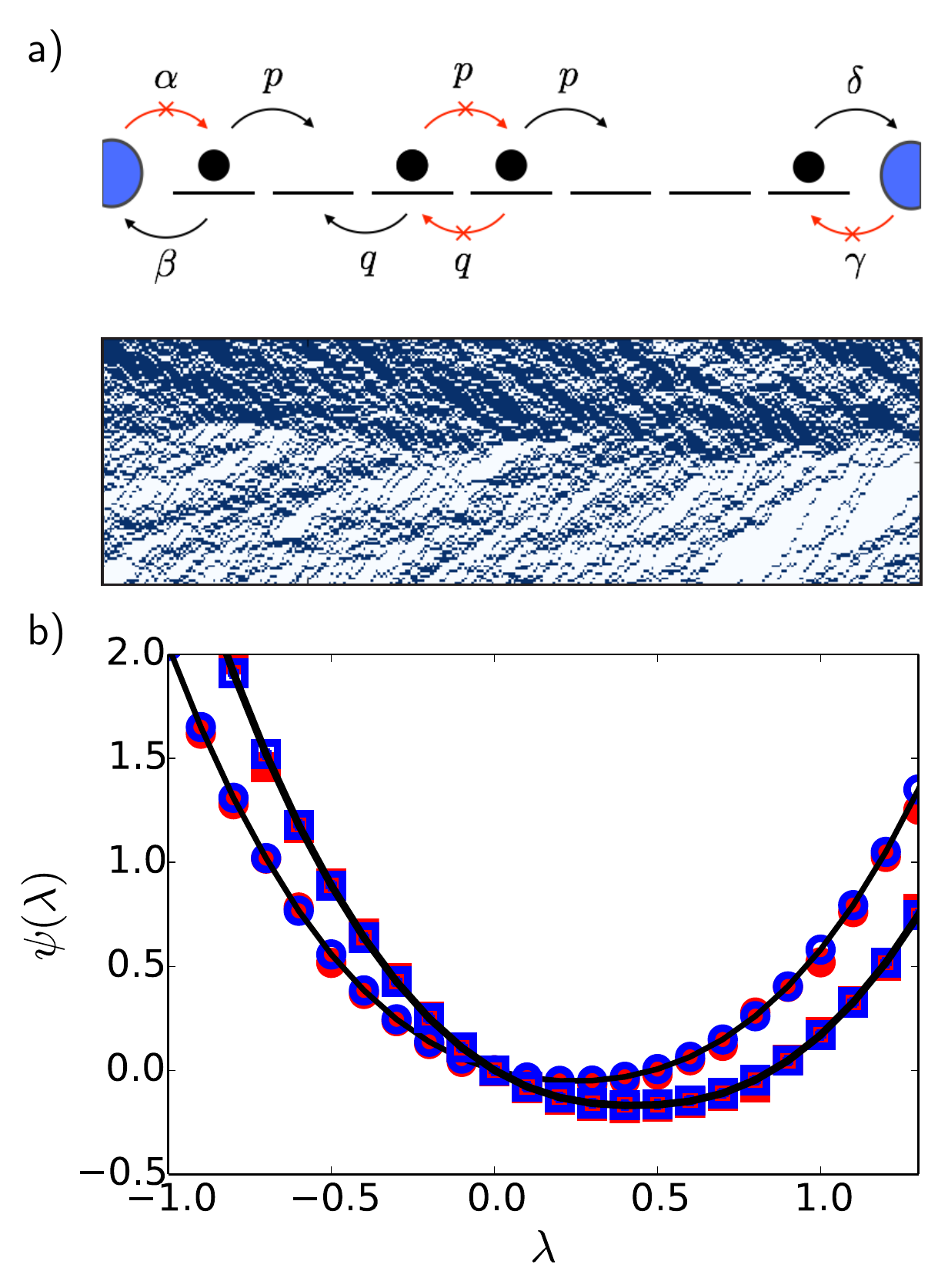}
\caption{Large deviation functions for the particle current in the SEP models. a) Schematic of the SEP model, 1$d$ lattice gas with nearest neighbor hopping rates and double occupancy exclusion. A typical trajectory shown in space-time, where the dark lattice sites denote an occupied position. b) Large deviation function for $L=8$ boundary (circles) and bulk (squares) driven SEP models, computed with TPS (red), DMC (blue) and exact diagonalization (solid line).}
\label{Fi:5}
\end{center} 
\end{figure}
To explore the origin of statistical uncertainties in the estimation of large deviation functions, we consider a family of models known as simple exclusion processes (SEP). These models are well studied examples of one dimensional transport and are considered paradigmatic models of nonequilibrium many-body physics\cite{schutz1998integrable}. The simple exclusion process evolves on a lattice with $L$ sites, and in the following we consider the open case whereby particles are injected or removed at site 0 and $L-1$. At a given time, $t$, the configuration of the system is defined by a set of occupation numbers, $n_i=\{0 \,1\}$, e.g. $\mathcal{C}(t)=\{0,1,..,1,1\}$. The probability of observing a given configuration at a specified time, $\rho(\mathcal{C}_t)$, can be found from the solution of the continuity equation in Eq. \ref{Eq:Master}, with a Markovian rate matrix, $\mathcal{W}$, whose elements have the form 
\begin{equation}
\mathcal{W}(\mathcal{C},\mathcal{C}') = -r(\mathcal{C}) + \sum_{\mathcal{C}\ne\mathcal{C}'} w(\mathcal{C},\mathcal{C}')
\end{equation}
where $r(\mathcal{C})=\sum_{\mathcal{C}} w(\mathcal{C},\mathcal{C}')$ is the exit rate and $w(\mathcal{C},\mathcal{C}')$ has the elements $w = w_L + \sum_i w_i + w_R$, where
$$
w_L =  \begin{bmatrix}
-\alpha& \gamma\\
\alpha& -\gamma
\end{bmatrix} \quad \quad 
\quad \quad w_R =\begin{bmatrix}
-\delta& \beta\\
\delta& -\beta
\end{bmatrix}
$$
$$
w_i = \begin{bmatrix}
0 & 0 & 0 & 0\\
0 & -q & p & 0\\
0 & q & -p & 0\\
0 & 0 & 0 & 0\\
\end{bmatrix}
$$
The matrices $w_L$ and $w_R$ act on the first and last sites, respectively, in a basis of single particle occupancy, i.e. $n_i = \{0 \,\, 1\}$. Particles are inserted at the boundaries with rates $\alpha$ and $\gamma$, and removed with rates $\beta$ and $\delta$. The matrices $w_i$ act on all $L$ sites, and are expressed in a multi-particle basis, i.e. $n_i n_{i+1} = \{00 \, \, 01 \,\, 10 \,\, 11\}$. Particles move to the right with rate $p$ and to the left with rate $q$, subject to the constraint of single site occupancy. This hard core constraint results in correlations between particles moving on the lattice. These rates and the hard core exclusion are illustrated in Fig.~\ref{Fi:5}a.   

For most values of $\alpha,\gamma,\delta,\beta,q,$ and $p$, an average current of particles flows through the system. We denote the particle current by $Q(\tobs)$, which is equal to the number of hops to the right minus the number of hops to the left over a time $\tobs$, 
\begin{equation}
Q(\tobs) = \sum_{t=1}^{\tobs} \sum_{i=0}^{L-1} \delta_{i+1}(t)\delta_i(t-1) - \delta_i(t) \delta_{i+1}(t-1)
\end{equation}
where $\delta_{i}$ is the Kronecker delta function for site $i$ and the sum runs over the lattice and $\tobs$. Finite steady state currents can be generated either by boundary driven or bulk driven processes. For $p=q$, boundary driven currents occur when $\alpha, \beta$, and $\gamma, \delta$ are such that they establish mean densities on the left and right ends of the lattice that are unequal. For $q\ne p$, bulk driven currents are produced, and remain finite in the infinite lattice size limit, unlike boundary driven currents which decay following Fick's law as $1/L$. For $q\ne p$ and different values of the boundary rates, the model exhibits a nontrivial phase diagram, with different density profiles and mean currents\cite{kolomeisky1998phase}. A representative trajectory in the maximum current phase\cite{kolomeisky1998phase} is shown in Fig.~\ref{Fi:5}a. As a relatively low dimensional many-body system, SEP models represent an ideal testing ground for studying statistical errors in the computation of large deviation functions.

We have computed large deviation functions for parameters that include both bulk and boundary driven transport. For the case of boundary driven transport, we study a model with $\alpha=\delta=0.9$, $\beta=\gamma=0.1$, and $q=p=0.5$. For the case of bulk driven transport, we study a model with $\alpha=\delta=\beta=\gamma=0.5$, and $p=0.75$ and $q=0.25$. For both cases we consider fairly small lattices, $L=8$. The dynamics are propagated by sampling the trotterized rate matrix using kinetic Monte Carlo\cite{bortz1975new}, with a timestep of 0.05 that is commensurate with single particle moves. For calculations using TPS we simulated trajectories of length $\tobs=20$, using forward shooting and shifting moves. Trajectories are harvested using the acceptance criterion given in Eq.~\ref{Eq:TPSaccep}, with the bias $\exp(-\lambda Q(\tobs))$. Shooting and shifting points were optimized to result in a 30$\%$ acceptance ratio and around 10$^5$-10$^6$ trajectories were harvested for each value of $\lambda$. Large deviation functions were solved self-consistently with MBAR as in Eq.~\ref{Eq:Wham}. For calculations using DMC, we used integration times, $\tint=0.5$, $\tobs =100$, and a constant walker population of $\Nw=$2000. Branching probabilities were computed from Eq.~\ref{Eq:weights2}, and $\psi(\lambda)$ was estimated from Eq.~\ref{Eq:DMCpsi2}.  

Figures~\ref{Fi:5}b  show the large deviation functions computed for fluctuations of $Q$. As before there is quantitative agreement between both TPS and DMC. Figure~\ref{Fi:5}b compares results for both bulk and boundary driven cases obtained by sampling with TPS and DMC and exact numerical results obtained by computing the maximum eigenvalue of the many-body rate matrix using  exact diagonalization. Across the range of $\lambda$'s, the sampled values agree with those numerically exact results. In all cases, the simple convex shape of the large deviation functions are a consequence of the fluctuation theorem, this one derived by Lebowitz and Spohn for Markov jump processes\cite{lebowitz1999gallavotti}. For a 1d system such as SEP, the total entropy production along a trajectory is linearly proportional to the current\cite{esposito2010three}. The proportionality constant is known as the affinity, $A=\ln (\alpha \delta p^{L-1}/\beta \gamma q^{L-1})/L+1$, and results in a slightly different symmetry for current fluctuations, 
\begin{equation}
\psi(\lambda) = \psi(A-\lambda)
\end{equation}
where for an equilibrium system, $A=0$, and the fluctuation theorem is again a statement of microscopic reversibility\cite{lebowitz1999gallavotti}. For SEP models, the value of $A$ depends on whether the current is generated by bulk or boundary driving, and as a consequence, the curves in Fig~\ref{Fi:5}b are symmetric about different values.

\begin{figure}
\begin{center}
\includegraphics[width=8.5cm]{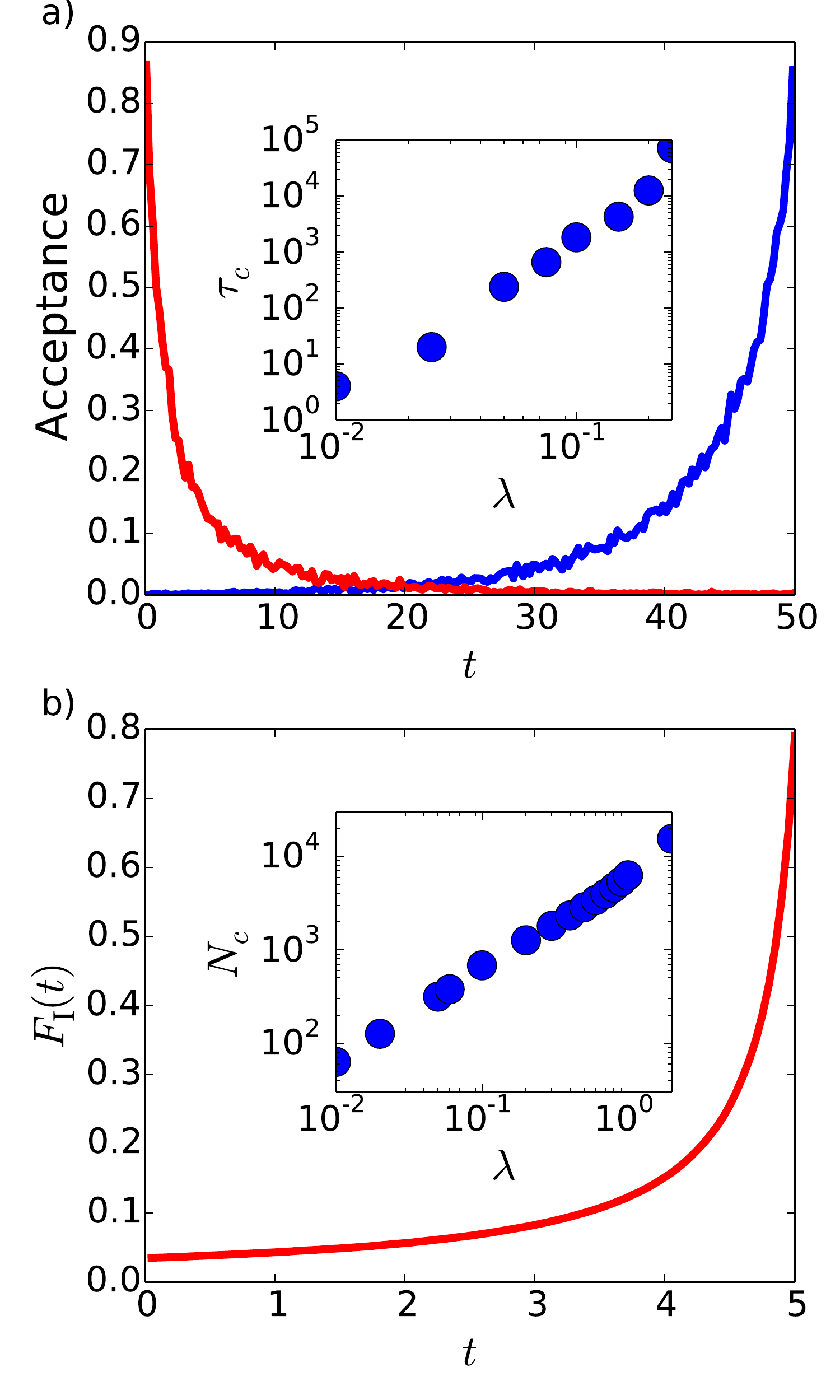} \\
\caption{Statistical efficiencies of large deviation function calculations for the SEP model. a) The main figure shows the average acceptance ratio computed with TPS for $\tobs=50$  and $\lambda=0.1$. The red line is computed for forward shifting moves and the blue line is computed for forward shooting moves. The inset is the correlation time in number of TPS moves as a function of $\lambda$. b) The fraction of independent walkers computed with DMC for  $\tint=0.5$ and $\lambda=0.1$. The inset is the number of correlated walks as a function of $\lambda$ for the same system.}
\label{Fi:SEPerr}
\end{center} 
\end{figure}

 As before the main systematic errors in converging $\psi(\lambda)$ scale as $1/\tobs$ for TPS and $1/\Nw$ for DMC. For the systems considered here with $L=8$, for $|\lambda| \le 1$ the large deviation functions can be converged. However with increasing $\lambda$ the statistical effort grows significantly. For the TPS algorithm, the origin of the statistical inefficiencies can be understood by computing the average acceptance probability for moves made at different times along the trajectory. Shown in Fig.~\ref{Fi:SEPerr}a is the average acceptance probability for the boundary driven SEP model, computed for $\lambda=0.1$. We find that the acceptance probability is large at the ends of the trajectory and vanishing in its middle. This confinement of acceptance illustrates that at this value of $\lambda$ only small perturbations to a trajectory can be accepted. The decay of acceptance probability away from the ends becomes increasingly abrupt for larger $\lambda$.

The efficiency of a Monte Carlo algorithm depends on how many independent samples can be generated with a given computational effort.
Diffusion in trajectory space will be small if only small changes to the trajectory are made at each Monte Carlo step. In order to quantify how correlated trajectories are during the sequential updating of TPS, we compute a correlation function of current fluctuations. For trajectories separated by $\tau$ TPS moves, we define the trajectory space function as,
 \begin{equation}
I(\tau) =  \langle \delta Q[\mathscr{C}_0]  \delta Q[\mathscr{C}_\tau] \rangle 
\end{equation}
where $\delta Q$ is trajectory's current minus its ensemble average, $Q - \langle Q \rangle$. This correlation function is monotonically decreasing, so it is uniquely characterized by its decay time which we define by.
\begin{equation}
\tau_c =  \int_0^\infty d\tau \, I(\tau)/I(0) \, .
\end{equation}
The correlation time $\tau_c$ is a time measured in TPS Monte Carlo moves. The inset of Fig.~\ref{Fi:SEPerr}a shows $\tau_c$ as a function of $\lambda$. The correlation time has been computed using moves that result in an acceptance probability of 30\%, so that we are effectively probing the characteristic decay of acceptance probability off of the ends trajectory. Because this results in only small changes to a trajectory to be accepted, subsequent trajectories stay correlated over more moves for increasing $\lambda$. We find that over a small range of $\lambda=0.01-0.2$ this increase is abrupt, growing over 4 orders of magnitude. To have accurate estimates, many statistically independent samples must be generated by a Monte Carlo procedure but this rapidly decreasing statistical efficiency prohibits convergence for $\lambda>1$. As the bias being offset by the acceptance criteria is an exponential of $\lambda$ times an extensive order parameter, we expect that the correlation time is a exponentially sensitive function of $\lambda$. Because the order parameter grows with system size or trajectory length, the range of possible updates to the trajectory becomes small for large systems or $\tobs$, making studying all but the smallest fluctuations for large systems cumbersome, if not numerically impossible.  

A similar analysis can be done to understand the statistical efficiencies of DMC. In the cloning algorithm, rather than shifting the bias into an acceptance criteria as TPS does, it funnels probability into a population dynamics. Correlations between trajectories result from walkers that have common ancestors. Recently, Nemoto et al. introduced a useful measure of this correlation, as the fraction of independent clones, $F_\mathrm{i}(t)$ which measures the probability that a descendant of a walker at a time $t$  survives up to the final time $\tobs$\cite{Takahiro2016}. This quantity is shown in Fig.~\ref{Fi:SEPerr}b along a $\tint=0.5$ step for the bulk driven SEP model with $\Nw=$2000.  Just as for the acceptance ratio in TPS, $F_\mathrm{i}(t)$ is strongly peaked at $\tobs$ and decays quickly from the boundary.  Analogously, $F_\mathrm{i}(t)$ reports on the number of independent walkers and the decay time from $\tobs$ sets a rate over which walkers become correlated. 

We can quantify the statistical efficiency of DMC by computing the total number of correlated walkers, $N_c$, over the repeated iterations of propagation and population dynamics. In terms of $F_\mathrm{i}(t)$, 
\begin{equation}
N_c = \Nw (1 - F_\mathrm{i}(0)) 
\end{equation}
where, $F_\mathrm{i}(t)$, is evaluated in the limit of the time going to $0$ where we generically find that it plateaus. This estimate is the number of clones that share a common ancestor somewhere along the simulation. This quantity is plotted in the inset of Fig.~\ref{Fi:SEPerr}b as a function of $\lambda$. We find that the number of correlated walkers is strongly increasing as a function of $\lambda$. This means that as we probe rarer fluctuations, fewer walkers carry significant weight. In fact, to represent the large deviation function an exponentially large number of walkers as a function of $\lambda$ must be used. As the statistical errors are inversely proportional to the number of independent walkers\cite{nemoto2016finite}, the estimates for large $\lambda$ become exponentially worse. 

For both algorithms, the origin of sampling difficulties is in the exponentially decreasing overlap between the ensemble generated by the bare dynamics, $P[\mathscr{C}(t_N)]$ and the ensemble that contributes most to the large deviation function at a given $\lambda$'s through the acceptance criteria in TPS or the branching weights in DMC, $P_\lambda[\mathscr{C}(t_N)] $. Both algorithms propagate trajectories without any information from the exponential basis that distinguishes these distributions, and attempt to offset all of the exponential weight onto a subsequent importance sampling step. In the case of TPS, all of the weight is put on the acceptance criteria for making moves in trajectory space. This results in an acceptance rate that becomes exponentially small, and subsequently trajectories that become correlated over exponentially large number of moves. In the case of DMC, all of the weight is put into the branching step. This results in  a walker population that must be exponentially large in order to overcome the exponentially different walker weights. Because the targeted ensemble is weighted by a factor that is an exponential of a time and system size extensive quantity, these deficiencies severely limit the current practical utility of these standard algorithms.

%
%

\begin{figure*}[t]
\begin{center}
\includegraphics[width=17cm]{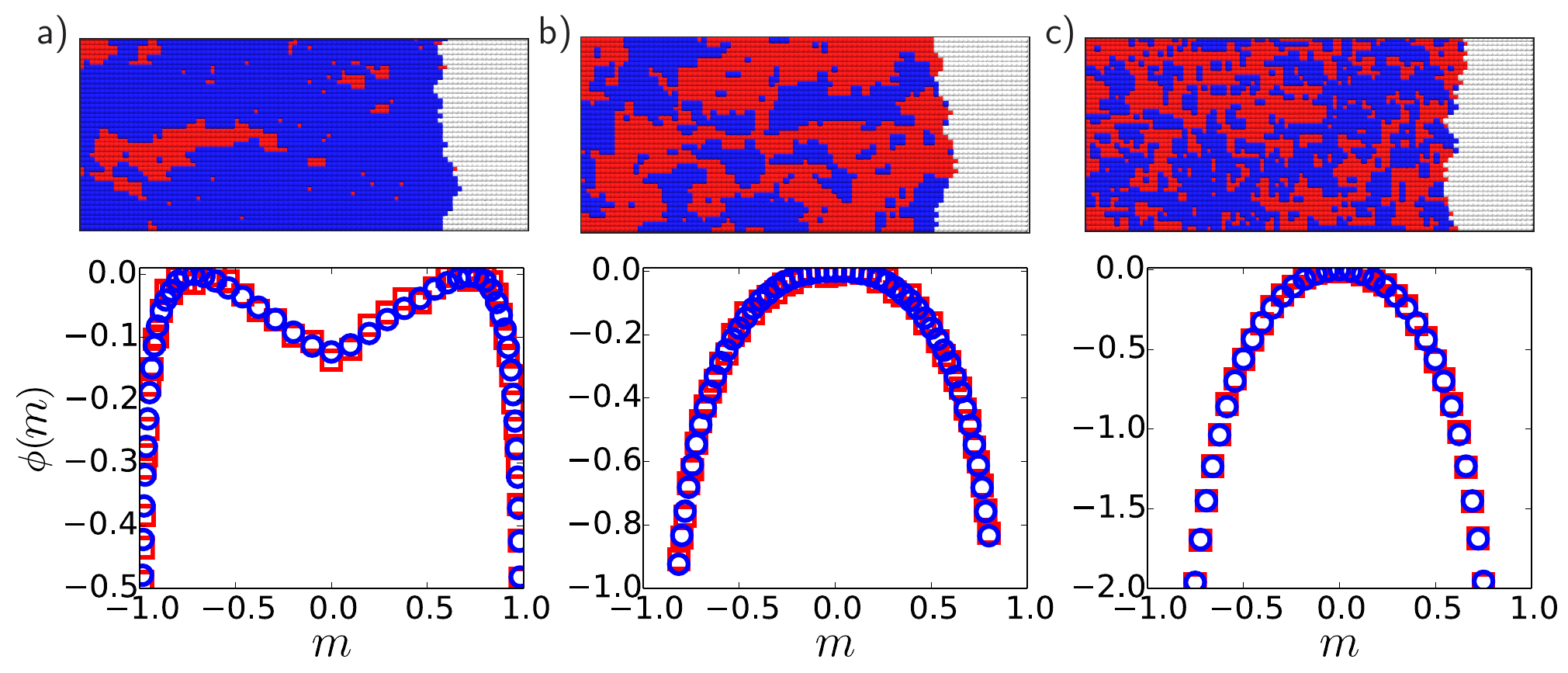}
\caption{Rare fluctuations of the net magnetization for the MRSOS model computed with TPS (red) and DMC (blue) for $\mu=$ a) -0.5, b) 0.0 and c) 0.5. The top panels show representative configurations of the lattice computed with these parameters.}
\label{Fi:rsos}
\end{center} 
\end{figure*}

\subsection{Alternate biasing form}
The final system we consider is the two dimensional, multicomponent, restricted solid-on-solid model (MRSOS)\cite{jullien1985scaling}. This model was recently studied in the context of nonequilibrium crystal growth, where it was observed that by tuning the growth rate the system can undergo an order-disorder transition with concomitant critical point and regions of bistability\cite{whitelam2014self}. As a two dimensional model and one exhibiting critical fluctuations, it presents an ideal testing ground for exploring sampling difficulties that result from large typical fluctuations and strong finite size effects. Also, because it has the potential for bistability, over short times the rate function may exhibit non-convex parts, analogous to equilibrium free energies for finite size systems at phase coexistence. Indeed few importance sampling techniques have been able to importance sample these fluctuations efficiently. A notable exception is recent work by Klympko et al., who derived an effective dynamics capable of sampling rare trajectories for this model\cite{Klymko2017rare}.

The model is defined by an energy function, 
\begin{equation}\label{Eq:Ising}
E = -J \sum_{<i,j>} s_i s_j - \mu \sum_i | s_i | 
\end{equation}
where $s_i =\{\pm 1, 0\}$,  $J>0$ is the standard ferroelectric coupling, $\mu$ is the chemical potential for adding a spin of either sign, and $<i,j>$ denotes a sum over only nearest neighbors.  The model is evolved with a Metropolis Monte Carlo dynamics, with uniform addition and deletion probabilities. The spin additions or deletions are constrained to only a thin layer at the interface of the growing lattice. Specifically, we impose the solid-on-solid restriction\cite{kim1989growth} that does not allow for the addition or removal of a spin with $s = \pm 1$ to a site if the difference in the new height and the height of adjacent sites with $s = \pm 1$ exceeds 1. This essentially freezes the interior of the lattice, allowing for spin flips and deletions only at the interface. Starting from an initial condition with $s_i = 0$ for all $i$, and a flat interface at the origin, for $\mu>-0.5$, on average the lattice will grow through the sequential additions of spins, creating a net mass current.  Periodic boundary conditions are imposed in the direction parallel to the interface. In the direction perpendicular to the interface, mixed boundaries are imposed with a fixed boundary at the origin and an open boundary in the direction of net growth.\\

The energy function in Eq.~\ref{Eq:Ising} is just the Ising Hamiltonian and at equilibrium the structure should belong to a two dimensional system independent of growth mechanism. Consequently, in a quasi-static growth limit, we expect the system to follow the physics of the two dimensional Ising model with a phase transition occurring at an inverse temperature $J/k_\mathrm{B} T = \ln(1+\sqrt{2})/2 \sim 0.44$. At $J/k_BT = 0.75$ the system is deep within the ordered state so that the expectation is that structure should be demixed, with large domains of spins of one sign. Remarkably the system undergoes a dynamical phase transition dependent on the growth rate as determined by the chemical potential, from a demixed state in the quasi-static limit, to a mixed state at high growth rates\cite{whitelam2014self}. Example configurations on either side of the transition are shown in Fig. \ref{Fi:rsos}. For fixed $J$, and small $\mu=-0.5$, typical configurations are large domains of mostly spin up or down with equal probability. At $\mu=$0.0, domains of spin up and spin down of many sizes span the system. For large $\mu=0.5$, the fast growth results in completely mixed lattice of spin up and spin down, with no long ranged correlations. Because of the kinetic constraint, these nonequilibrium patterns cannot easily anneal.

Depending on which side of the dynamical transition the system is on, the different spatial correlations will be reflected in the distribution of magnetization-per-particle, 
\begin{equation}
M(\tobs)=\sum_{i=1}^{N(\tobs)} s_i 
\end{equation}
where $N(\tobs)$ is the number of added spins $s_i = \pm 1$. At long times the system relaxes to a nonequilibrium steady state with fixed growth rate $N(t) \propto \tobs$. In order to compute the distribution function for $m=M(\tobs)/N(\tobs)$ in the long time limit, we could bias ensembles of trajectories in TPS or DMC with a field conjugate to $m$ as was done in the previous sections. However, because of the potential bistability of $p(m)$, such a linear bias would not help sampling near $ m =0$ when $\langle |m| \rangle $ is finite. Instead, we choose to bias the system with a factor of the form,
\begin{equation}
\label{Eq:harmonic}
p_{\kappa,M^*}(m) = p(m) e^{-\kappa (M-M^*)^2-\tobs \psi(\kappa,M^*)}
\end{equation}
where the biased distribution is now a function of two parameters, $\kappa$ and $M^*$, as is the normalization constant $\psi(\kappa,M^*)$. In the limit that $\kappa$ is large and positive, trajectories will be localized near $M^*$, allowing us to sample arbitrary magnetizations, independent of the geometry of the underlying distribution. A consequence of such a bias is that $\psi(\kappa,M^*)$ is no longer a cumulant generating function, but inverting Eq. \ref{Eq:harmonic} 
\begin{equation}
\label{Eq:umbrellaWHAM}
\phi(m) = \phi_{\kappa,M^*}(m) - \kappa (M-M^*)^2/\tobs -  \psi(\kappa,M^*)
\end{equation}
we still can relate the biased rate function to the unbiased one, where $ \phi(m) =- \ln p(m)/\tobs$. Such a bias changes the acceptance criteria in TPS to
\begin{equation}
\mathrm{Acc}[ \mathscr{C}_\mathrm{o}\rightarrow \mathscr{C}_\mathrm{n}] = \mathrm{Min}\left ( 1,e^{-\Omega}e^{-\eta[\mathscr{C}_\mathrm{n},\kappa,M^*]+\eta[\mathscr{C}_\mathrm{o},\kappa,M^*])} \right )\\
\end{equation}
where $\Omega$ is the path entropy that depends on the type of MC move and
\begin{equation}
\eta[\mathscr{C},\kappa,M^*] = \kappa (M[\mathscr{C}]-M^*)^2
\end{equation}
and $M[\mathscr{C}]$ is the time integrated value of the magnetization for the trajectory $\mathscr{C}$.  

Due to the quadratic bias, the branching probability for DMC has to be altered so that the long time limit adheres to the required functional form of the weight. With a linear bias, the weight is simply multiplicative, so the branching rate can be easily computed. In the non-linear case this is no longer possible and we must resort to the telescopic form, which requires additional book keeping. For harmonic bias, the DMC branching probability is given by
\begin{equation}
k_i(t) = \frac{\exp [-\kappa (\Delta_i[\mathcal{C}_{0},\mathcal{C}_{t}] - \Delta_i[\mathcal{C}_{0},\mathcal{C}_{t-\tint}] )]}{\sum_j^{N_\mathrm{w}}\exp [-\kappa (\Delta_j[\mathcal{C}_{0},\mathcal{C}_{t}] - \Delta_j[\mathcal{C}_{0},\mathcal{C}_{t-\tint}])}
\end{equation}
where
\begin{equation}
\Delta_i[\mathcal{C}_{0},\mathcal{C}_{t}] = (M_i[\mathcal{C}_{0}\rightarrow \mathcal{C}_{t}]-M^*)^2
\end{equation}
is the harmonic bias up to time $t$. In both TPS and DMC, provided $\kappa$ and $M^*$ are chosen such that different ensembles of trajectories overlap, we can use histogram reweighting together with the relation in Eq. \ref{Eq:umbrellaWHAM} to reconstruct $\phi(m)$ deep into the tails of the distribution.

Shown in Fig.~\ref{Fi:rsos} are TPS and DMC results for the MRSOS model using harmonic biases. Calculations were accomplished for $\mu=-0.5,0.0,0.5$ and $J=0.75$ using system sizes that were approximately $L=40$ and $\tobs=200$. The TPS calculations were run with $\kappa=2$ and 40 windows, with $M^*/\tobs$ chosen to cover $m=-1,1$, and trajectories of 200 MC steps sampled with shifting and shooting moves. In all, 10$^5$ trajectories were collected for each $\mu$. The DMC calculations were run using $\Nw=10000$ and $\tint=1$. For all cases a spring constant, $\kappa=2$ was used and 40 values of $M^*$ were spread uniformly across the range of $m=-1,1$. 

For all cases studied, the DMC and TPS results are statistically identical.  For large $\mu=0.5$ as expected, the system exhibits a single minimum, centered at $m=0$, with gaussian fluctuations about the average. At intermediate driving, $\mu=0$, the system exhibits one broad minimum, with fluctuations that go like $\phi(m) \propto m^6$ for small $m$. The concomitant susceptibility computed under such conditions is divergent in the large system, long time limit, as expected for a critical transition. For small $\mu=-0.5$,  we find the system exhibits two minima, each stable basin centered at finite values of $m$ that are equal and opposite as required by symmetry\cite{nguyen2016design}. If the non-convexity of $\phi(m)$ is related to a surface tension in spacetime, we would expect that in the limit of large $\tobs$ and large $L$ that the rate function becomes flat about $m=0$. We indeed find this to be the case, and further find that depending on the ratio of $\tobs$ and $L$,  interfaces can preferentially form in the smaller of the spatial or temporal directions, yielding a second metastable minimum about $m=0$, consistent with recent calculations\cite{Klymko2017rare}.

While  generalizing the calculations beyond linear biases allows TPS and DMC to sample non-convex rate functions, and study a complex model of self-assembly, the system sizes and times  are still restrictive.  Indeed without addition importance sampling, performing finite size scaling to determine the critical exponents for the transition is prohibitively expensive. 

\section{Conclusion}

We have shown how large deviation functions within nonequilibrium steady states can be computed with trajectory based importance sampling.  In particular, we have studied the systematic and statistical errors associated with two Monte Carlo algorithms, TPS and DMC, to estimate large deviation functions. We found that though they are implemented in different ways, the two algorithms sample the same ensemble of trajectories and suffer from related difficulties. 

The unifying feature of these two algorithms is that they augment the propagation of the bare system dynamics with \emph{a posteriori} importance sampling. In the case of TPS, this is done with a Metropolis acceptance step after proposing a new trajectory from an existing one. 
%
In the case of DMC, the bare dynamics are augmented with a population dynamics, whereby walkers are cloned or deleted after short trajectories are generated. In both cases, the proposal step of the Monte Carlo chain is done without any guidance from the rare events that contribute to the large deviation function. As a result, for exponentially rarer fluctuations, the Monte Carlo chain requires exponentially more samples of the targeted stationary distribution as the overlap between it and the proposed distribution becomes exponentially small. This means that in practice low dimensional systems, or small fluctuations, can be systematically converged, but rare fluctuations of high dimensional systems are impractical to compute. We have illustrated this with a number of simple models where exact numerical calculations can be done, on and off lattice. By making simple adjustments to the algorithm and using standard histogram reweighting tools we could push the algorithm to study the MRSOS model. This model exhibits a complex phase diagram, including a critical point. However, systematic studies on the finite size scaling properties are outside the reach of these current algorithms.

These results highlight a potential avenue for further development, where guiding functions are incorporated into the proposal step, in analogy to what is done in quantum diffusion Monte Carlo\cite{Reynolds1982}. To do this in the context of nonequilibrium steady states, requires developing auxiliary dynamics that can push the trajectories to the rare fluctuations that contribute to the large deviation function. In principle, for Markovian dynamics an optimal auxiliary process always exists and can be constructed by a canonical transformation known as the Doob transform\cite{chetrite2015nonequilibrium}. However, to construct such a process exactly requires finding the large deviation function. In Part 2 of this work, we construct approximations to the Doob transform and use those as guiding functions for Monte Carlo sampling\cite{ray2017importance2}, and show that
  orders of magnitude in statistical efficiency can be gained by relatively simple approximate guiding functions. 

{\bf Acknowledgements:} D.T.L was supported by UC Berkeley College of Chemistry. U. R. was supported by the Simons Collaboration on the Many-Electron Problem and the California Institute of Technology. G. K.-L. C. is a Simons
Investigator in Theoretical Physics and was supported by the California Institute of Technology.

%

\end{document}